\documentstyle[psfig,mncite,subfigure]{mn}

\begin{document}

\title{A search for starlight reflected from HD 75289 b}
\author
[C.Leigh, A.Collier~Cameron, S.Udry, J-F.Donati, K.Horne, D.James and A.Penny]
{\large Christopher Leigh$^{1}$, Andrew Collier Cameron$^{1}$, Stephane Udry$^{2}$, Jean-Fran\c{c}ois Donati$^{3}$, Keith Horne$^{1}$, \\
\large David James$^{4}$ and Alan Penny$^{5}$ \\
\small $^{1}$ University of St Andrews, St Andrews, Fife, KY16 9SS, U.K \\
$^{2}$ Observatoire de Gen\`{e}ve, 51 ch. des Maillettes, 1290 Sauverny, Switzerland \\
$^{3}$ Laboratoire d'Astrophysique Observatoire Midi-Pyr\'{e}n\'{e}es, 14 avenue Edouard Belin, F-31400 Toulouse, France \\ 
$^{4}$ Observatoire de Grenoble, F-38041 Grenoble, Cedex 9, France \\
$^{5}$ Rutherford Appleton Laboratory, Chilton, Didcot, Oxon, OX11 0QX, U.K}
\date{\today}

\maketitle

\begin{abstract}
We have used a doppler tomographic analysis to conduct a deep search for the starlight
reflected from the planetary companion to HD 75289. In 4 nights on VLT2/UVES in January
2003, we obtained 684 high resolution \'{e}chelle spectra with a total integration time of
26 hours. We establish an upper limit on the planet's geometric albedo $p<0.12$ (to the
99.9\% significance level) at the most probable orbital inclination $i\simeq 60^\circ$,
assuming a grey albedo, a Venus-like phase function and a planetary radius $R_{p}=1.6
R_{Jup}$. We are able to rule out some combinations of the predicted planetary radius and
atmospheric albedo models with high, reflective cloud decks.
\newline
\\
{\bf Key Words:} Planets: extra-solar - Planets: atmosphere - Stars: HD 75289
\\
\\
\\
\end{abstract}

\section{Introduction}
\label{sec:intro}

The existence of a close planetary companion to the G0V star HD75289 was first reported
by \scite{udry2000}. In common with 19 of the 117 extrasolar planets known, HD75289b is
found to orbit within 0.1 AU of its parent star. Confirmation of the gas-giant nature
of the close-orbiting exoplanet, HD209458b \cite{charb2000,henry2000b}, suggests that
similar objects may reflect enough starlight to allow a direct planetary detection.
Spectral models of these ``Pegasi planets'' \cite{barman02,sudarsky03} show that
scattered starlight dominates over thermal emission at optical wavelengths.  
\scite{sudarsky03} suggest the high effective temperature and relatively low surface
gravity of HD75289b may favour the formation of relatively bright, high-altitude
silicate cloud decks, which act to scatter starlight back into space before being
absorbed by alkali metals in the deeper atmosphere.

Here we report the results of observations, conducted over 4 nights in January 2003 on
VLT2/UVES \'{e}chelle spectrograph, aimed at detecting the starlight reflected from
HD75289b. In Section~\ref{sec:parameters} we summarise the basic physics which
underpins our analysis, whilst sections~\ref{sec:obser} and \ref{sec:process} review
the acquisition, reduction and processing of the raw \'{e}chelle spectra, using techniques
described more comprehensively in \scite{cameron02} and \scite{leigh03}. Finally, in
section~\ref{sec:results} we use our results to place upper limits on the grey geometric 
albedo of the planet.

\section{System Parameters}
\label{sec:parameters}

HD75289 (HIP43177) is a G0 main sequence star with parameters listed in
Table~\ref{parameters}. High-precision radial-velocity (RV) measurements obtained
between Nov 1998 and Oct 1999 \cite{udry2000} were used to identify a planetary
companion HD75289b, whose properties (as determined directly from RV studies or inferred
using the estimated stellar parameters) are also summarised in Table~\ref{parameters}.
Since the planet publication \cite{udry2000}, the orbital solution has been updated with
new Coralie RV measurements using a weighted cross-correlation scheme \cite{pepe02} with
an appropriate template.

As the planet orbits its host star, some of the starlight incident upon its surface is
reflected towards us, producing a potentially detectable signature within the observed
spectra of the star. This signature takes the form of faint copies of the stellar absorption
lines, Doppler shifted due to the planet's orbital motion and greatly reduced in intensity
due to the small fraction of starlight the planet intercepts and reflects back into space.

With our knowledge of the stellar mass and the planet's orbital period we can estimate
the orbital velocity of the planet $V_{p}$ (see \pcite{cameron02}).
The apparent RV amplitude $K_{p}$ of the reflected light is given by
\begin{equation}
K_{p} = V_{p} \sin i = 147 \sin i~~{\rm km~s^{-1}} ,
\label{eq:kp}
\end{equation}
where the orbital inclination $i$ is, according to the usual convention, the
angle between the orbital angular momentum vector and our line of sight.
 
For all but the lowest inclinations, the orbital velocity amplitude of the planet is
much greater than the broadened widths of the stellar absorption lines. Hence, lines in
the reflected-light spectrum of the planet should be Doppler shifted well clear of their
stellar counterparts, allowing a clean spectral separation for most of the orbit.

By isolating the reflected planetary signature, we are in effect observing the
planet-star flux ratio ($\epsilon$) as a function of orbital phase ($\phi$)
and wavelength ($\lambda$), where
\begin{equation}
\epsilon(\alpha,\lambda)\equiv\frac{f_{p}(\alpha,\lambda)}{f_{\star}(\lambda)}
        =p(\lambda)g(\alpha,\lambda)\frac{R_{p}^{2}}{a^{2}} .
\label{eq:fluxratio}
\end{equation}
The phase function $g(\alpha,\lambda)$ describes the variation in the
planet-star flux ratio with the phase angle $\alpha$, here $\alpha$
is the angle subtended at the planet by the star and the observer,
and varies as $\cos\alpha = - \sin i \cos\phi$. As the
shape of $g(\alpha,\lambda)$ is unknown, current practice is to adopt
a specific phase function in order to express the results in terms of
the planet-star flux ratio that would be seen at phase angle zero
\begin{equation}
\epsilon_{0}(\lambda)=p(\lambda)\frac{R_{p}^{2}}{a^{2}} ,
\label{eq:eps0}
\end{equation}
where $p(\lambda)$ is the wavelength-dependent geometric albedo and the orbital
distance $a$ is constrained by Kepler's third law. For reasons described in
\scite{cameron02} and \scite{leigh03}, the $g(\alpha,\lambda)$ we adopt is a polynomial
approximation to the empirically determined phase function for Venus \cite{hilton92}.

\begin{table}
\begin{center}
\begin{tabular}{cc}
$\bf{Star:}$ \\
Spectral Type           		& G0V $^{1}$ \\
$m_{v}$                 		& 6.35 (0.013) $^{2,3}$ \\
Distance (pc)           		& 28.9 (0.46) $^{2}$ \\
$T_{\rm Eff}$ (K)           		& 6135 (40) $^{4}$ \\
$M_{*} (M_\odot)$       		& 1.22 (0.05) $^{5,6,7}$ \\
$R_{*} (R_\odot)$      			& 1.25 (0.05) $^{5,6,7}$ \\
$\left[{\rm Fe/H}\right]$     		& 0.27 (0.06) $^{4}$ \\
$P_{rot}$ (d)               		& 16.0 (3.0) $^{8,9}$ \\
v$\sin i ({\rm km~s^{-1}})$   		& 3.8 (0.6) $^{10}$ \\
Age (Gyr)               		& 5.6(1.0) $^{8,11}$ \\
\\
$\bf{Planet:}$ \\
Orbital Period $P_{orb}$ (d) 		& 3.5091 (0.0001) $^{10}$ \\
Transit Epoch $T_{0}$ (JD)      	& 52651.1566 (0.1251) $^{10}$ \\
$K_{*} ({\rm m~s^{-1}})$          	& 53.4 (0.6) $^{10}$ \\
a (AU)                          	& 0.0483 (0.0020) $^{10}$ \\
$M_{p}\sin i (M_{Jup})$         	& 0.455 (0.028) $^{10}$ \\
\end{tabular}
\end{center}
\caption[]{System parameters (and uncertainties) for HD75289 and its planetary companion. 
$^{1}$  \scite{gratton89}, $^{2}$ \scite{esa97}, $^{3}$ \scite{gray01}, $^{4}$ \scite{santos01},
$^{5}$  \scite{fuhrmann98},$^{6}$ \scite{gonzalez2000}, $^{7}$ \scite{gonzalez01}, $^{8}$ \scite{udry2000},
$^{9}$ \scite{noyes84}, $^{10}$ \scite{udry2000} revised for this paper, $^{11}$ \scite{donahue93}.
}
\label{parameters}
\end{table}

In the event of a planet detection, analysis of the data should allow us to
determine two fundamental properties:

(i) $K_{p}$, the planet's projected orbital velocity, from which we obtain the system 
inclination and planet mass, since $M_{p}\sin i$ is known from the star's Doppler
wobble,

(ii) $\epsilon_{0}$, the maximum flux ratio observed where, by making
theoretical assumptions about the nature of the planet, we can invoke
equation (\ref{eq:eps0}) to constrain $R_{p}$ or $p$.

\section{Observations}
\label{sec:obser}

\begin{table}
\begin{tabular}{ccccc}
\\
UTC start             & Phase   &  UTC End  & Phase  & $N_{exp}$  \\
                      &         &           &        &          \\
Jan 14 01:00:04 & 0.396 & Jan 14 09:29:44 & 0.497 &  188 \\
Jan 15 00:56:42 & 0.680 & Jan 15 09:21:32 & 0.780 &  173 \\
Jan 21 02:40:18 & 0.410 & Jan 21 09:34:54 & 0.492 &  183 \\
Jan 22 03:35:30 & 0.706 & Jan 22 09:41:01 & 0.778 &  140 \\
\\
\end{tabular}
\caption{Journal of observations. The UTC mid-times and orbital phases are shown
for the first and last spectral exposures secured on each night of observation in
January 2003. The number of exposures is given in the final column.}
\label{tab:journal}
\end{table}

We observed HD75289 during Jan 2003 using the blue arm of the UV-Visual \'{E}chelle
Spectrograph (UVES), mounted on the 8.2-m VLT/UT2 (Kueyen) telescope at the Paranal
Observatory in Chile. The detector was a single EEV CCD-44 array windowed to an image
format of $2048\times 3000$ 15.0 $\mu$m pixels, and centred at 475.8 nm in order 102 of
the 41.6 g mm$^{-1}$ \'{e}chelle grating, giving good wavelength coverage from 406.5
nm to 522.0 nm. For reasons described in greater detail in \scite{cameron02}, we
believe the prominence of Rayleigh scattering, combined with the high density of
absorption lines within this spectral range, offer the best chance of detection.
With an average pixel spacing close to 1.5 $\rm km~s^{-1}$, the 
FWHM intensity in the central orders of the thorium-argon calibration spectra was
found to be 4.5 pixels, suggesting an effective resolving power $R=43000$.

Table \ref{tab:journal} details the 4 nights of observations that contribute to this
analysis. Each of the stellar spectra were exposed between 100s and 300s, depending upon
the seeing, in order to expose the CCD to $\sim$ 40000 ADU pix$^{-1}$ in the brightest
parts of the image. A typical exposure in average seeing (0.8 arcsec) yielded $\sim 4
\times 10^{5}$ electrons per pixel step in wavelength in the brightest orders after
extraction. The CCD's 7s fast-readout time allowed an observing efficiency $>$ 90\%,
and saw a gain of 2.4 and readout noise of $\sim$ 6 $e^{-}$.

\section{Data Processing}
\label{sec:process}

\begin{figure*}
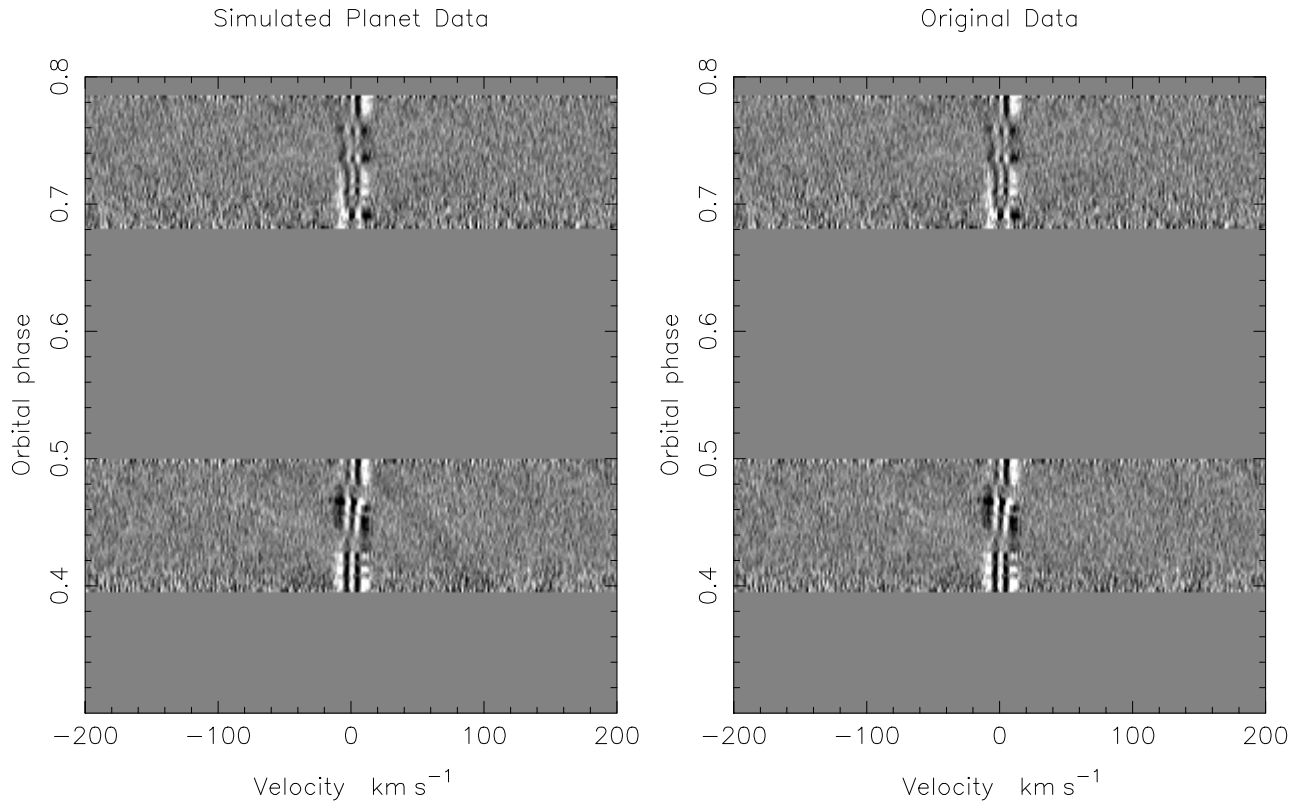

\begin{center}
\begin{tabular}{cc}
\psfig{figure=MD964L_fig1.ps,bbllx=70pt,bblly=70pt,bburx=400pt,bbury=490pt,width=8.2cm} &
\psfig{figure=MD964L_fig2.ps,bbllx=70pt,bblly=70pt,bburx=400pt,bbury=490pt,width=8.2cm}
\end{tabular}
\end{center}
\caption[]{Time series of deconvolved profiles derived from the VLT/UVES spectra, secured
over 4 nights in Jan 2003. The left plot (a) demonstrates the result of adding a 
simulated planet signal at an inclination of $80^\circ$, assuming grey geometric
albedo $p=0.4$ and a radius, $1.6~R_{Jup}$. The planetary signature appears as a dark
sinusoidal feature crossing from right to left as phase increases and centred on the superior
conjunction at phase 0.5. The right plot (b) shows the original data, where no such
trail is evident. The greyscale runs from black at $-10^{-4}$ times the mean stellar
continuum level, to white at $+10^{-4}$. The velocity scale is in the reference frame of 
the star. 
}
\label{fig:ph_noplanet}
\end{figure*}

One-dimensional spectra were extracted from the raw images using an   
automated pipeline reduction system built around the Starlink ECHOMOP
and FIGARO packages.

After initial tracing of the \'{e}chelle orders, an extraction process subtracted the bias
from each frame, cropped it, determined the form and location of the stellar profile on each
image, subtracted a linear fit to the scattered-light + sky background across the spatial
profile, and performed an optimal (profile and inverse variance-weighted) extraction of
orders across the full spatial extent of the object+sky region. In all, 25 orders (Nos
90-114) were extracted from each image, giving good spectral coverage from 406.5 to 522.0
nm. Following extraction, the S:N in the continuum of the brightest orders was typically 650
per pixel step.

Flat-field balance factors were applied prior to extraction by summing the 50-100
flat-fields taken at the start and end of each night. Although the UVES instrument
was very stable between observations, we found that the noise was slightly reduced
($\sim$ 2\%) by the use of nightly flat-fields rather than master flat-fields
combining all 4 nights' data.

\subsection{Extracting the planet signal}
\label{sec:aldecon}

\begin{figure*}
\begin{center}
\begin{tabular}{cc}
\psfig{figure=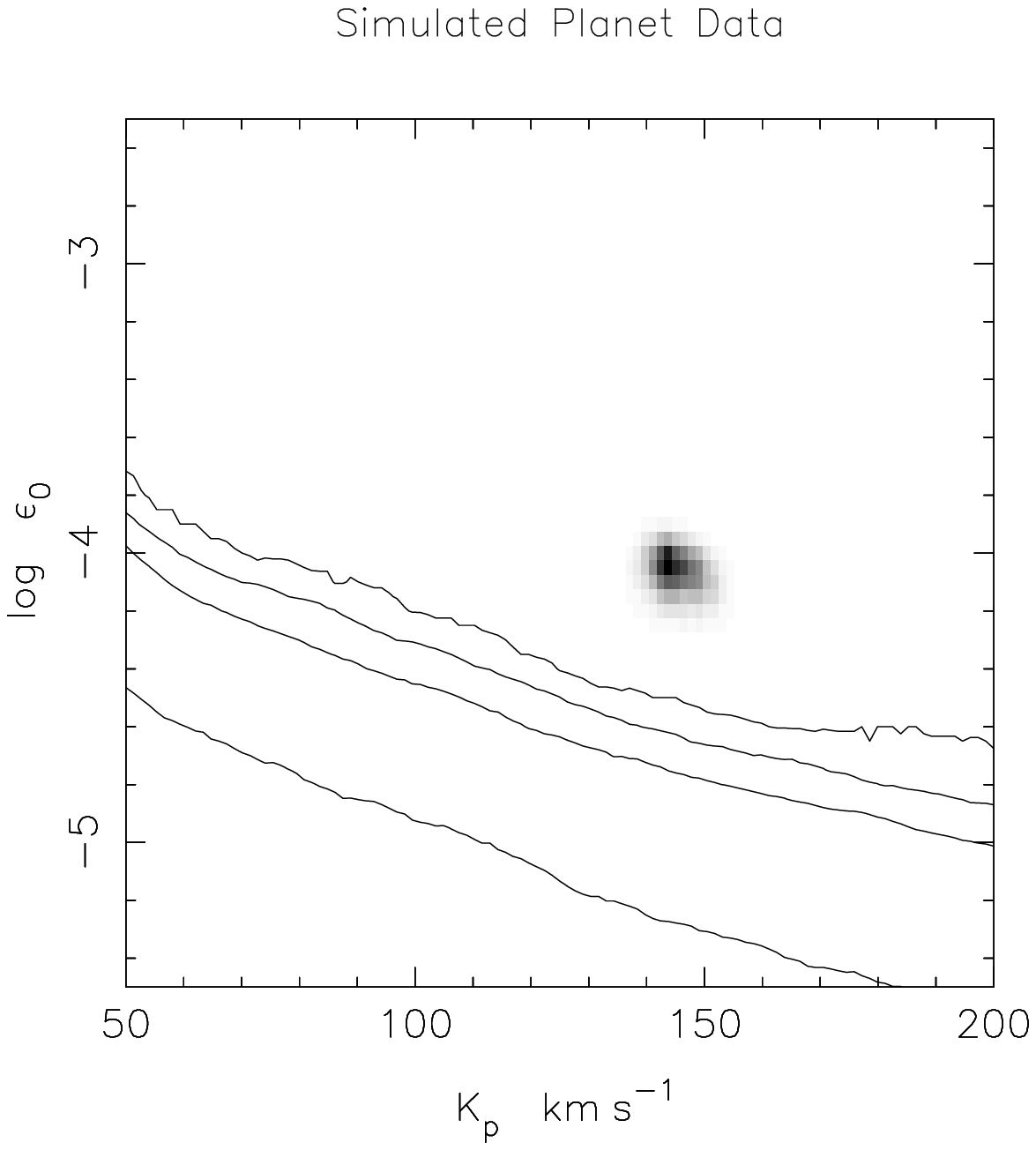,bbllx=110pt,bblly=70pt,bburx=475pt,bbury=420pt,angle=-90,width=8.6cm} &
\psfig{figure=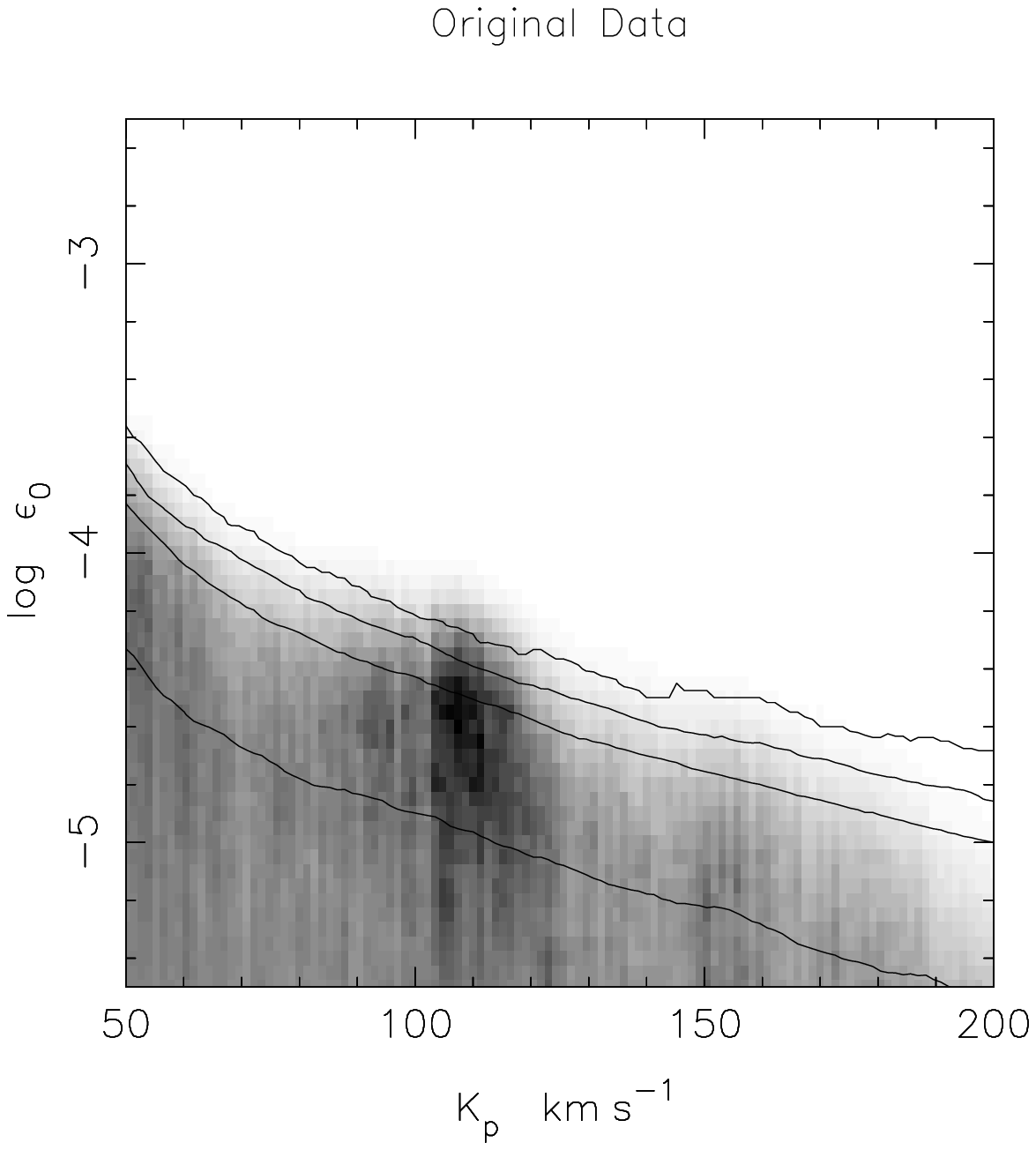,bbllx=110pt,bblly=70pt,bburx=475pt,bbury=420pt,angle=-90,width=8.6cm}
\end{tabular}
\end{center}
\caption[]{Relative probability map of model parameters $K_{p}$ and
$\log \epsilon_{0} = \log p(R_{p}/a)^{2}$, derived from the VLT/UVES observations of
HD75289. The greyscale denotes the probability relative to the best-fit model, increasing
from 0 for white to 1 for black. The contours show the confidence levels at which
spurious detections due to non-Gaussian noise can be ruled out. From top to bottom, they
show the 99.9\%, 99.0\%, 95.4\%\ and 68.3\%\ confidence limits on $\epsilon_{0}$ for
fixed $K_{p}$. The left plot (a)
demonstrates how the simulated planet signal is recovered through the analysis with
ease, and is detected well above the 99.9\%\ confidence limit. The right plot (b)
containing the original data, however, demonstrates little evidence for a
planetary detection. It is important to note that the contours only give the 
probability of a false detection if the value of $K_{p}$ is known in advance, which is 
not the case here. The slightly increased probability density seen close to $K_{p} = 110
~{\rm km~s^{-1}}$ (within the $K_{p}$ distribution predicted by \pcite{leigh03b})
returns a large false-alarm probability of 32\% when considering all plausible
values $35 < K_{p} < 147~{\rm km~s^{-1}}$, which is far too uncertain to claim as genuine.}
\label{fig:limits}
\end{figure*}

The first step in extracting the reflected component is to subtract the direct stellar
component from the observed spectrum. The planet signal should then consist of faint
Doppler-shifted copies of each of the stellar absorption lines, at this stage buried deeply
in the noise. A detailed description of this procedure is given in \scite{cameron02}
Appendix A. After cleaning any correlated fixed-pattern noise remaining in the difference
spectra (see \scite{cameron02} Appendix B), we create a composite residual line profile by
deconvolving each residual spectra with a list of stellar absorption line strengths
(\scite{cameron02} Appendix C). The deconvolution results in a 32-fold improvement in the
S:N, by combining the properties of 2360 images of 1744 absorption lines listed within the
observed wavelength range. The composite line profiles are then stacked by phase to display
temporal variations in brightness and RV, as at Fig.~\ref{fig:ph_noplanet}.
Any planet signal present  appears as a dark sinusoidal feature in velocity
space centred on the superior conjunction at phase 0.5.

The central pattern of distortions in Fig.~\ref{fig:ph_noplanet} is consistent with
sub-pixel shifts in the position of the stellar spectra with respect to the detector
over the course of the night. Fortunately they only affect a range of velocities at
which the planet signature would in any case be indistinguishable from that of the
star.

\subsection{Simulated planet signatures}
\label{sec:simulated}

We verified that a planetary signal is preserved through the above sequence of operations,
in the presence of realistic noise levels, by adding a simulated planetary signal to the
observed spectra (Fig.~\ref{fig:ph_noplanet}a). The fake signal also acts to calibrate 
the strength of any genuine observations and the associated confidence contours. 
Detailed explanations of the simulated planet calibration can be found in 
\scite{cameron02} and \scite{leigh03}. To ensure a strong signal we used a simulated
planet of radius 1.6 ${R}_{Jup}$ and wavelength-independent geometric albedo $p=0.4$,
which when viewed at zero phase angle should give a planet-star flux ratio
$\epsilon_{0}=f_{p}/f_{*}=9.71\times 10^{-5}$.

Except for the case of a tidally locked system, any relative motion between the surface of the
planet and the surface of the star will result in a difference between the line widths of the
incident and reflected spectra. From the known stellar parameters (Table~\ref{parameters}), we
have conducted Monte Carlo trials to estimate that the reflected component within the HD75289
spectra will exhibit typical line widths close to $13~{\rm km~s^{-1}}$. In order to 
correctly
mimic the reflected starlight we have selected the faster rotating G0V star HD1461 as our
planetary template. A star which is of very similar temperature and elemental abundance to
HD75289.

\subsection{Matched Filter Analysis}
\label{sec:aldecon}

The next step involved a matched-filter analysis, described by \scite{cameron02}
Appendix D, to search for features in the time-series of composite residual profiles
whose temporal variations in brightness and RV resemble those of the expected 
reflected-light signature.

The relative probabilities of the $\chi^{2}$ fits to the data for different
values of the free parameters $R_{p}/a$ and $K_{p}$ are conveniently displayed
in greyscale form as a function of $K_{p}$ and $\log \epsilon_{0} =\log
p(R_{p}/a)^{2}$. Fig.~\ref{fig:limits} details the probability map for the
simulated (a) and original (b) observations, normalised to the $\Delta \chi^{2}$
of the best-fit model.

To set an upper limit on the strength of any planet signal, or to assess the likelihood
that any signal detection is spurious, we need to compute the probability of obtaining
similar improvements in $\chi^{2}$ by chance alone. We achieve this using a
``bootstrap'' procedure to construct empirical distributions for confidence testing,
using the data themselves. In 3000 trials, we randomized the order in which the 684
observations were secured, but associated them with the original sequence of dates and
times. Genuine signals are scrambled in phase, but re-ordered data are as capable
as the original data of producing spurious detections through chance alignments of
systematic errors along a sinusoidal path through the data. We record the least-squares
estimates of $\log \epsilon_{0}$ and the associated $\chi^{2}$ as functions of $K_{p}$
in each trial.

The percentage points of the resulting bootstrap distribution are shown as
contours in Fig.~\ref{fig:limits}. From bottom to top, the contours represent the
68.4\%, 95.4\%, 99.0\% and 99.9\% bootstrap upper limits on the strength
of the planet signal. Thus the 99.9\% contour represents the value of 
$\log \epsilon_{0}$ that was only exceeded in 3 of the 3000 trials at each
$K_{p}$.

\section{Results}
\label{sec:results}

The results of this analysis appear in the form of a relative probability map of model
parameters $K_{p}$ and $\log \epsilon_{0} =\log p(R_{p}/a)^{2}$, shown at
Fig.~\ref{fig:limits}. The calibrated confidence contours allow us to apply constraints
to the geometric albedo of the planet, given certain theoretical assumptions about the
system.

\subsection{Upper Limits on Grey Albedo}
\label{sec:grey_results}

\begin{figure}
\psfig{figure=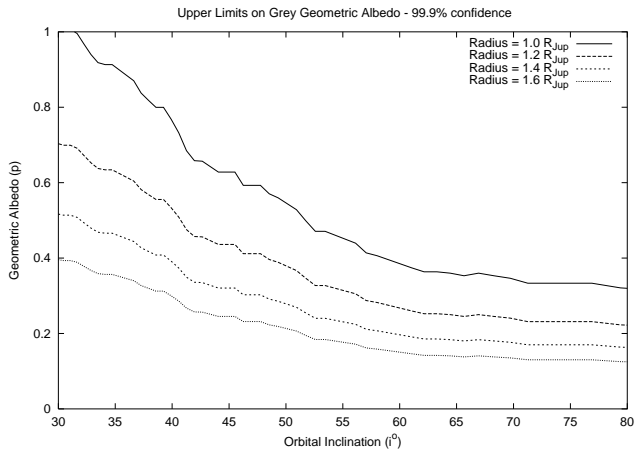,angle=-90,width=8.6cm}
\caption[]{Upper limits (99.9\% confidence) on the geometric albedo $p$ as a
function of inclination and radius, assuming
the atmosphere of the planet imposes no wavelength dependence on reflected light,
i.e. a grey albedo model. The contours, from top to bottom, represent upper
limits assuming theoretical radii for the orbiting planet of 1.0, 1.2, 1.4 and
1.6 $R_{Jup}$, covering orbital inclinations $30^{\circ}<i<80^{\circ}$.
If the planet exhibits the radius, 1.6 $R_{Jup}$, and inclination, $i=75^{\circ}$,
predicted by \scite{leigh03b}, then our observations suggest we can be 99.9\%
confident of a grey geometric albedo $p<0.12$.}
\label{fig:grey}
\end{figure}

The grey albedo model makes the unlikely assumption that the planet-star flux ratio remains
independent of wavelength. By adopting a theoretical radius for the planet we can use 
equation (\ref{eq:eps0}) to constrain its geometric albedo. Fig.~\ref{fig:grey}
details the upper limits on the albedo at the 99.9\% level of significance, for a
range of possible inclinations and radii.

We find that the albedo limits are less strongly constrained with low inclination systems.
This is a natural consequence of the matched-filter analysis, where fitting such models
incorporate more pixels close to the noisier ripples in the central stellar region of the
deconvolved profile.

If we adopt the \scite{leigh03b} theoretical radius for HD75289b of $R_{p} = 1.6~R_{Jup}$, 
we find the planet-star flux ratio, at the most probable $K_{p} \simeq
127~{\rm km~s^{-1}}$, is constrained to $f_{p}/f_{*} < 4.18 \times 10^{-5}$ at the 99.9\%
level, i.e. a geometric albedo less than 0.12. This would strongly rule out atmospheric
models which incorporate high-altitude, reflective cloud decks such as the Class V model
of \scite{sudarsky2000}.

\vspace{-3mm}

\section{Conclusion}
\label{sec:conclusion}

We have observed HD75289 over 4 nights in Jan 2003, using the VLT2/UVES instrument, in an
attempt to detect starlight reflected by the known close-orbiting planetary companion. The
excellent stability of the spectrograph, combined with the reasonable observing conditions,
have enabled us to produce deep upper limits on the geometric albedo of the planet. In
truth, with our knowledge of the system and current theoretical predictions, we would have
expected a reasonably unambiguous detection of the planet. However, we find very little 
evidence to suggest its presence. Possible reasons for the discrepancy are,

(i) The inclination of the system is such that the spectral features of the planet do 
not cleanly separate from the stellar features. i.e. maximum Doppler shift is not
sufficient to disentangle the planets signal. However, evidence generated \cite{leigh03b} 
using the known system parameters, suggests an inclination $i \gg 20^{\circ}$.

(ii) Our choice of phase function $g(\alpha,\lambda)$ is wrong. It may be that
close-orbiting Pegasi planets are less prone to back-scatter incoming starlight
than we see with Jupiter or Venus \cite{hovenier89,seager2000}.

(iii) Although the transiting planet HD209458 has indicated the presence of a gas giant,
it may be that HD75289b is a more compact terrestrial planet reflecting substantially
less starlight (see \pcite{guillot96}).

(iv) It could be that the geometric albedo is inherently low (cf. Class IV CEGP of 
\pcite{sudarsky2000}), with starlight absorbed deep in the atmosphere of a planet
where no high-level clouds are present to reflect the incident light. However, at the
wavelengths observed, we do expect a more significant contribution from Rayleigh 
scattering.

Whatever the cause, these observations provide a strong test for developing theoretical
models which aim to predict the atmospheric nature of these objects. This is a field in
desperate need of continued observations on the brightest of these Pegasi planets, as 
are detailed in the work of \scite{leigh03b}.

\vspace{-0.2cm}

\bibliography{MD964L_ref} \bibliographystyle{mn}

\vspace{-0.4cm}

\section*{Acknowledgements}

This work is based on observations with the VLT/UT2 (Kueyen) telescope, operated by the
European South Observatory at the Paranal Observatory on Cerro Paranal in the Atacama
Desert, Chile. All data processing was conducted using Starlink Project supported
hardware and software.

ACC and KDH were supported by PPARC Senior Fellowships and CJL by a PPARC Postgraduate
Studentship during the course of this work.

\end{document}